\input{psfig}
\input{epsf}

\def\mev{\,{\rm Me\kern-0.1em V}}
\def\gev{\,{\rm Ge\kern-0.1em V}}

\documentstyle[11pt]{article}

\renewcommand{\baselinestretch}{1.8}
\begin{document}
\vspace*{-1.25in}
\small{
\begin{flushright}
FERMILAB-PUB-98/181-T \\[-.1in] 
June, 1998 \\
\end{flushright}}
\vspace*{.75in}
\begin{center}
{\Large{\bf  An Efficient Algorithm for  QCD \\
with Light Dynamical Quarks }}\\
\vspace*{.45in}
{\large{A.~Duncan$^1$, 
E.~Eichten$^2$,  and
H.~Thacker$^3$}} \\ 
\vspace*{.15in}
$^1$Dept. of Physics and Astronomy, Univ. of Pittsburgh, 
Pittsburgh, PA 15260\\
$^2$Fermilab, P.O. Box 500, Batavia, IL 60510 \\
$^3$Dept.of Physics, University of Virginia, Charlottesville, 
VA 22901
\end{center}
\vspace*{.3in}
\begin{abstract}
A new approach to the inclusion of virtual quark effects in lattice 
QCD simulations is presented. Infrared modes which build in the
chiral physics in the light quark mass limit are included exactly
and in a gauge invariant way. At fixed physical volume the number of 
relevant infrared modes does not increase as the continuum limit is 
approached. The acceptance of our procedure does not decrease substantially in the
limit of small quark masses.
Two alternative approaches are discussed for including systematically
the remaining ultraviolet modes. In particular, we present evidence that
these modes are accurately described by an effective action involving
only small Wilson loops.
\end{abstract}

\newpage
\renewcommand{\baselinestretch}{1.0}

\section{Introduction}

Recent studies of the origin and role of exceptional configurations \cite{mqa,qed2,qnchart}
leading to extremely noisy hadron correlators in quenched lattice
QCD for light quark masses have underlined the importance of nonlocal
topological fluctuations in determining the chiral physics of the
quenched theory. It has been known for a long time \cite{coleman} that such fluctuations
completely alter the behavior of the theory in the light quark mass
limit once the quark determinant is included in a full dynamical
calculation. Moreover, the appearance  of spurious real modes \cite{mqa} of the
Wilson-Dirac operator on finite lattices in association with such
fluctuations accounts for the increasing frequency of exceptional 
configurations at stronger coupling and low quark mass. These connections
all suggest that reliable calculations in the chiral limit of lattice QCD
require an accurate treatment of the low eigenmodes of the quark Dirac
operator. 

Singularities of quark propagators in the quenched theory are automatically
regulated by corresponding zeroes of the quark determinant. However, this
regularization is only effective if the low eigenmodes of the quark Dirac
operator are treated precisely: in particular, valence and sea quark masses
must be identical. On the other hand, the high eigenmodes
(corresponding to imaginary masses for scales well above the QCD scale up
to the lattice cutoff) when integrated out  contribute to an effective
gauge-invariant gluonic action which, for physics on small momentum scales, simply amounts
to a redefinition of the scale of the theory. To the extent that the 
ground state hadron spectrum involves hadronic bound states with constituent quarks
all off-shell on the order of $\Lambda_{\rm QCD}$, it therefore seems likely
that high eigenmodes are simply irrelevant for spectrum calculations, even in full
QCD. By writing the fermion determinant in terms of the
hermitian operator $H\equiv \gamma_{5}(D\!\!\!/(A)-m)$ \cite{hermitian}
we are able to deal with a completely real spectrum. Moreover, 
 the individual eigenvalues have a direct physical interpretation as
a gauge-invariant measure of off-shellness of the quark fields (to see this, we recall that
for the free continuum theory, the eigenvalues of $H$ are simply $\pm\sqrt{p^{2}+m^{2}}$
for a quark mode of Euclidean momentum $p$). 

   We shall argue in this paper
that a separation of low and high eigenmodes
can in fact be carried out in a practical way in unquenched
lattice QCD calculations, leading to an efficient way of building in the 
important physics of the quark determinant in the chiral limit. Such a separation also
corresponds to a completely gauge-invariant and smooth interpolation between 
the quenched and full dynamics of the theory.  The procedure we propose also yields
as a byproduct very detailed and useful information about the infrared spectrum of the Dirac
operator which is known to be intimately related to the chiral physics of the theory
\cite{LeutSmil}, and central to the overlap formulation \cite{overlap} of lattice QCD.

In Section 2 we describe a regularized version of the fermionic determinant
which interpolates smoothly between the quenched and full theory, in a  way
which allows for the selective inclusion of fermionic modes in a predetermined
momentum range (typically from zero up to a given cutoff $\mu$). 
This regularization is amenable to an analytic perturbative
calculation in which the role of the high eigenmodes contributing to the
full fermionic determinant can be clearly isolated. Such a regularization
can be studied analytically in
abelian 4D gauge theory, where the $\mu$-dependence of the determinant 
for large $\mu$ and low momentum 
is seen to reduce to a shift of bare coupling (or in the lattice context,
of scale). 

 In Section 3  we  describe the results of some simulations in 2 dimensional lattice QED
(QED2), which has proven to be an extremely useful testbed for exploring features of the
Dirac-Wilson spectrum in lattice gauge theory. Here, and henceforth in all the
numerical simulations, we employ a regularization of the fermionic determinant in
terms of a sharp mode cutoff which is physically equivalent to the smooth regularization
of Section 2 but suitable for numerical implementation in large systems.
It is shown that the
fluctuations of the full fermionic determinant in an exact dynamical simulation of QED2 
are essentially restricted to a small fraction (for the lattices studied here,
essentially the lowest few percent)
of the spectrum. Comparisons of pseudoscalar correlators computed in the quenched and
full dynamical theory are made with an approximate simulation where only the 
lowest ten percent of the eigenvalues of the hermitian operator $\gamma_{5}(D\!\!\!\!/-m)$
are included in the fermionic determinant. The truncated determinant simulations
essentially reproduce the full dynamical results. Two characteristic features of unquenched
gauge theory, the suppression of topologically nontrivial sectors and the breaking of the
string due to shielding, are also illustrated using the truncated determinant 
approach in QED2. Of course, in the case of QED2, the superrenormalizability
of the theory implies that the high eigenmodes are basically inert, in distinction to 
the case in 4D gauge theory where these modes will necessarily introduce a further
logarithmic rescaling due to the variable screening effect of virtual quark/antiquark pairs
at different length scales.

 In Section 4 we describe in detail the algorithm we have employed for the simulations
of full QCD (on a 12$^3$x24 lattice at $\beta$=5.9 and inverse lattice spacing
$a^{-1}=$1.78 Gev for the quenched theory) with a truncated determinant. The Lanczos procedure
allows reliable extraction of Dirac eigenmodes up to energies $\sim$370 MeV, certainly enough to 
include the essential low energy chiral physics of QCD.   Moreover, the Lanczos procedure extracts
the needed small eigenvalues rapidly as the spectrum is relatively sparse there. Unlike the case of propagator inversion,
the Lanczos method is stable even in the presence of very small eigenvalues provided these are
not too dense. We also discuss
some aspects of the Monte Carlo 
dynamics (acceptance rate and equilibration time) for our update procedure, in which pure gauge
heat bath sweeps alternate with Metropolis accept/reject steps for the truncated determinant.
A crucial point is that we do not see a dramatic fall in the acceptance rate of our 
procedure as we go to lighter quark masses.

In Section 5 we present the results of our truncated determinant
simulations of QCD4. The initial study involves
runs on a 12$^3$x24 lattice at $\beta$=5.9 and at three kappa values (0.1570, 0.1587 and 0.1597), reaching in the lightest case a
pion mass on the order of 280 MeV. Pseudoscalar  meson masses are
measured and a value for the critical hopping parameter extracted.
The inclusion of 100 quark eigenmodes (all modes up to $\sim$370 MeV)
eliminates the necessity for considering quenched chiral logs \cite{chirlog} in the chiral extrapolations.
The topological charge distribution is measured for different
quark masses and compared with the quenched result. As expected, nonzero topological
charge is strongly suppressed in the light quark limit.  Measurements of the string tension  
reveal clearly a screening of the quark-antiquark potential from the virtual sea
quarks, although the lattice used is still too small to allow us to see the asymptotic
flattening expected at large distances.

In Section 6 we show that the  high momentum modes can be included in a precise
way by a combination of the truncated determinant and multiboson methods \cite{multiboson}.
The procedure suggested in this paper is in a sense exactly complementary to the 
multiboson approach of L\"{u}scher. The latter approach treats the high eigenmodes
of the Dirac operator very well, but necessarily introduces errors whenever small
eigenvalues are present. In the chiral limit such modes become frequent and in fact
dominate the chiral physics. Here we propose treating these modes as precisely as
possible. Another approach to the inclusion of the
high modes, a loop Ansatz for the short distance piece of 
the quark determinant, is also discussed in this final section. 
Such an Ansatz, involving relatively short 
Wilson loops (up to length 6), is shown to give a 
very accurate description of
the high end of the quark determinant. 
Finally, in Section 7 we summarize our conclusions.

\section{Truncated determinants in gauge theory}

The separation of low and high eigenmodes in the fermionic determinant can
be accomplished in an analytically convenient way by smoothly switching off
the higher eigenvalues above a sliding momentum scale $\mu$. Given a matrix
${\cal M}$ then ${\rm det}(\tanh({\cal M/\mu}))$ reduces to unity for $\mu$
much below the smallest eigenvalue of ${\cal M}$ while reproducing the full
determinant (up to an irrelevant multiplicative factor) for $\mu$ much above
the highest eigenvalue. For a gauge theory, defining the hermitian operator
$H\equiv \gamma_{5}(D\!\!\!/(A)-m)$, then the effective action obtained
from integrating out each flavor of fermion of mass $m$ can be regularized
as the logarithm of the smoothly truncated determinant
\begin{eqnarray}
\label{eq:defD}
{\cal D}(\mu)&\equiv& \frac{1}{2}{\rm tr}\ln(\tanh(\frac{H^{2}}{\mu^{2}})) \\
    &=& {\rm tr}\{ \ln(1-e^{-2H^{2}/\mu^{2}})-\ln(1+e^{-2H^{2}/\mu^{2}})\}
\end{eqnarray}

This definition allows an analytic calculation of the regularized determinant
in weak coupling perturbation theory, which describes the $\mu$-dependence
of ${\cal D}(\mu)$ for $\mu$ well above the QCD scale. The calculation can be
carried out for a nonabelian lattice regularized theory, but we shall illustrate   
the procedure here for the case of a continuum 4-dimensional abelian gauge       
theory. Note that

\begin{eqnarray}
\label{eq:Kdefs}
H^{2} &=& K_{0}+K_{1}+K_{2} \\
K_{0} &\equiv& \Box +m^{2} \\
K_{1} &\equiv& \{ -i\partial\!\!\!/, A\!\!/ \}  \\
K_{2} &\equiv& A_{\mu}A_{\mu}
\end{eqnarray}
To second order in weak coupling perturbation theory, we may compute
${\cal D}(\mu)$ by expanding to first order in $K_2$ and to second
order in $K_1$. The calculation is lengthy but straightforward (details
will be given elsewhere)- here we quote the result only. Expressed in
terms of momentum space fields, one finds
\begin{equation}
\label{eq:perttanh}
{\cal D} = \int\frac{d^{4}k}{(2\pi)^{4}}\beta(k^{2},m,\mu)A_{\mu}(k)(
k^{2}\delta_{\mu\nu}-k_{\mu}k_{\nu})A_{\nu}(-k)
\end{equation}
The contribution of the high modes can be studied by taking $\mu$ 
large compared with the quark mass $m$ and, in the nonabelian case,
with the QCD scale. Then these modes affect the low energy physics
(i.e for $k <<\mu$) through the low momentum limit of $\beta(k^2)$.
Explicit calculation gives
\begin{equation}
\beta(k^{2},m,\mu) \simeq -\frac{1}{24}\ln(\frac{\mu^{2}}{m^2})+O(k^{2})
\end{equation}
which exactly corresponds to the expected $\mu$-dependence
of the screening shift in the 
running coupling induced by virtual fermionic modes in the momentum range
up to $\mu$.

The decoupling of the high fermionic modes suggests that lattice QCD  calculations
performed at weak enough coupling should be insensitive to the fluctuations
induced by eigenvalues of the Dirac operator much above the QCD scale, except for
an overall shift in the scale of the theory induced by renormalizations of the
coefficients of the low dimension operators making up the effective pure
gauge action. In particular, dimensionless ratios of physical quantities
should fairly soon become insensitive to inclusion of higher modes in the
fermionic determinant. In a superrenormalizable theory like QED2, this insensitivity
should even be apparent in dimensionful quantities, as we do not 
have a logarithmic running of scale in this case. 

\section{Truncated Determinant Algorithm in QED2}

 Abelian gauge theory in 2 space-time dimensions (the massive Schwinger model)
has proven  to be a marvellously manageable testbed for exploring in 
detail \cite{SmitVink,qed2} the spectral properties of the Dirac-Wilson operator. The computational
expense of performing even exact update full dynamical simulations is relatively
slight, essentially full information on the spectrum can be obtained configuration
by configuration, and the system mimics, at least qualitatively, many of the 
topological and chiral properties of 4 dimensional QCD. This model also turns out
to be a very useful starting point for investigating the relative importance of
the infrared and ultraviolet ends of the Dirac spectrum in a full dynamical
lattice simulation. 

Although the calculation of all the eigenvalues of $H$, and hence of ${\cal D}(\mu)$
as defined in the previous section, is perfectly feasible for 2D QED, the restriction
of practical numerical techniques for the much larger matrices of 4D QCD to the
low-lying eigenvalues suggest the use of a simpler truncation of the determinant, in which
the lowest (in absolute magnitude) $N_{\lambda}$ positive and negative eigenvalues of $H$ are
included and all higher modes dropped. As we shall see in Section 5, precisely such a 
truncation scheme matches on exactly to a very accurate representation of the high end
of the determinant in terms of an effective loop action.  Labelling positive eigenvalues of
$H$ as $\eta_{n}$ and negative eigenvalues as $\zeta_{n}$ (where the index runs in the
 direction of increasing absolute magnitude) we define
\begin{eqnarray}
\label{eq:discdet}
{\cal D}(N_{\lambda}) &\equiv&\frac{1}{2}\sum_{n=1}^{N_{\lambda}}\ln(\eta_{n}^{2}\zeta_{n}^{2})  \\
\hat{\cal D}(N_{\lambda}) &\equiv&\frac{1}{2}\sum_{n=N_{\lambda}+1}^{D}\ln(\eta_{n}^{2}\zeta_{n}^{2})
\end{eqnarray}
where 2$D$ is the dimensionality of the discrete Wilson-Dirac matrix for the lattice
theory. An exact full dynamical simulation would include the full (log) determinant
${\cal D}(N_{\lambda})+\hat{\cal D}(N_{\lambda})$ in the effective gauge action. The extent to which the
low eigenvalues determine the physics of the unquenched theory can be examined by
comparing the fluctuations- in a dynamical simulation- of ${\cal D}(N_{\lambda})$ with those
of $\hat{\cal D}(N_{\lambda})$ for various choices of  $N_{\lambda} << D$. These fluctuations are shown
graphically in Fig.1, for 40 configurations generated in a full dynamical simulation
using an exact update algorithm. In Fig.2 the fluctuations are shown  for 40 configurations in a    
quenched  simulation. The lattice used was 10x10 at $\beta$=4.5 with a bare quark mass of
0.095. Evidently the fluctuations are essentially all confined to the low end of
the spectrum. In the quenched case the size of the fluctuations at the infrared end
is considerably larger than  for the dynamical configurations, as configurations with
small eigenvalues are suppressed once the determinant factor is included in the 
update procedure.  The appearance of such configurations is intimately related
to the exceptional configurations encountered in quenched calculations at strong 
coupling and/or small quark mass.
\begin{figure}
\psfig{figure=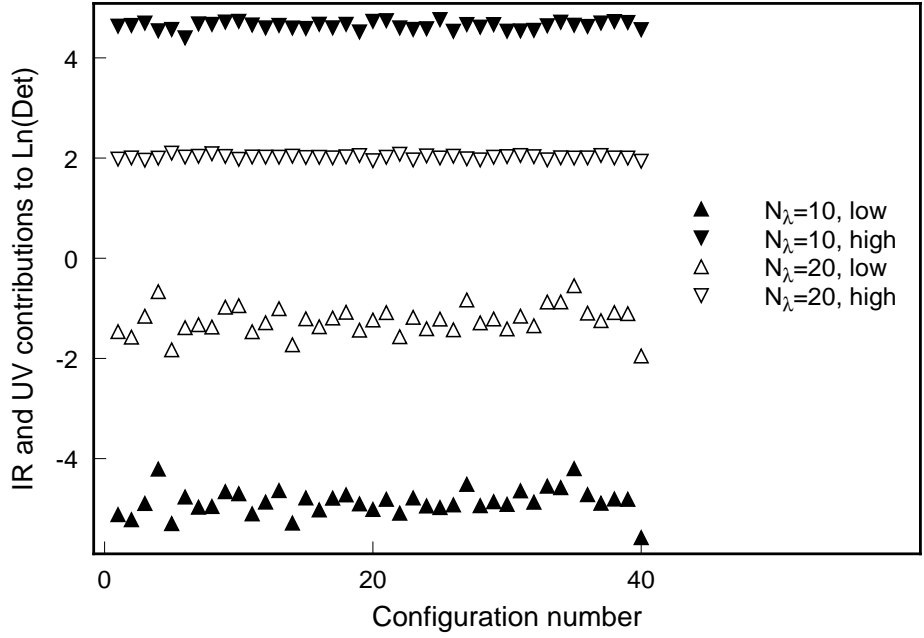,
width=0.95\hsize}
\caption{Fluctuations in ${\cal D}(N_{\lambda}),\hat{\cal D}(N_{\lambda})$, dynamical configurations.
 For ease of visibility, the curves have been shifted vertically.}
\label{fig:dynfluc}
\end{figure}

\begin{figure}
\psfig{figure=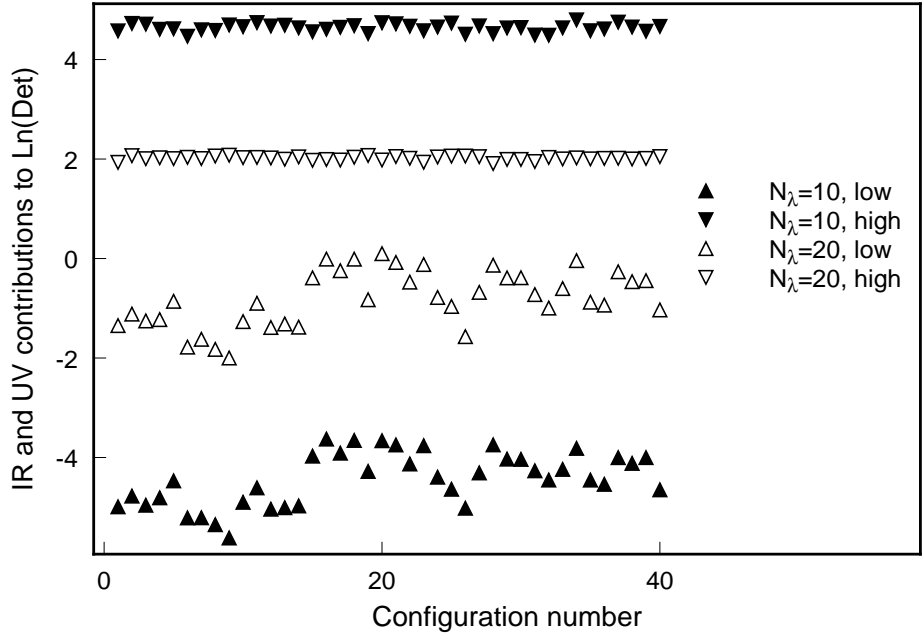,
width=0.95\hsize}
\caption{Fluctuations in ${\cal D}(N_{\lambda}),\hat{\cal D}(N_{\lambda})$, quenched  configurations}
\label{fig:qnchfluc}
\end{figure}

This behavior suggests an approximate unquenched algorithm in which 
only ${\cal D}(N_{\lambda})$ is used
in the determinant part of the effective lattice action. If $N_{\lambda}$ is chosen large
enough, all the nonperturbative infrared physics will be properly included. 
Exceptional configurations, in which there is an anomalously low eigenmode of the
Dirac-Wilson operator, are tamed in the expected way \cite{Lat97}, and the convergence
of the procedure can be examined simply by repeating the run for increasingly large $N_{\lambda}$. The update algorithm we have chosen is very simple: a number
(typically 5) of conventional
Monte Carlo  (Metropolis) sweeps are performed to obtain a new gauge configuration,
a new value for ${\cal D}(N_{\lambda})$ is calculated (for QED2 on a 10x10 lattice,
we can easily obtain all the eigenvalues by direct diagonalization) and compared
with that for the preceding configuration. Then the truncated determinant factor
is used to provide a Metropolis accept/reject criterion for the gauge configuration update. 
With $N_{\lambda}$=10,  the acceptance ratio was typically in the range of 50-75\%.
Measured quantities such as the pseudoscalar correlator decorrelated after a few
configuration updates (the statistical errors shown  include autocorrelation
times computed from the data). The results of such a procedure for the
pseudoscalar correlator are shown in Fig 3 for a quark mass (lattice units)
of 0.05. For such a small mass, using Wilson fermions, the quenched functional
integral is dominated by real pole contributions which appear in the simulation
as wildly noisy values for the correlators (one finds in a typical run of
800 sweeps values for the pion propagator at zero time separation ranging from
several thousand to zero, for a quantity averaging to order unity in the
dynamical theory). Instead we have plotted the quenched results
regularized by the pole shifting (or ``modified quenched approximation") procedure of
\cite{mqa}.  Evidently the full dynamical result for the pseudoscalar
correlator is reached with only a small fraction (in this case, about 10\%)
of the eigenmodes included in the 
fermionic determinant. When only 5\% of the eigenvalues are included, the result
is intermediate between the quenched and full dynamical values. 

\begin{figure}
\psfig{figure=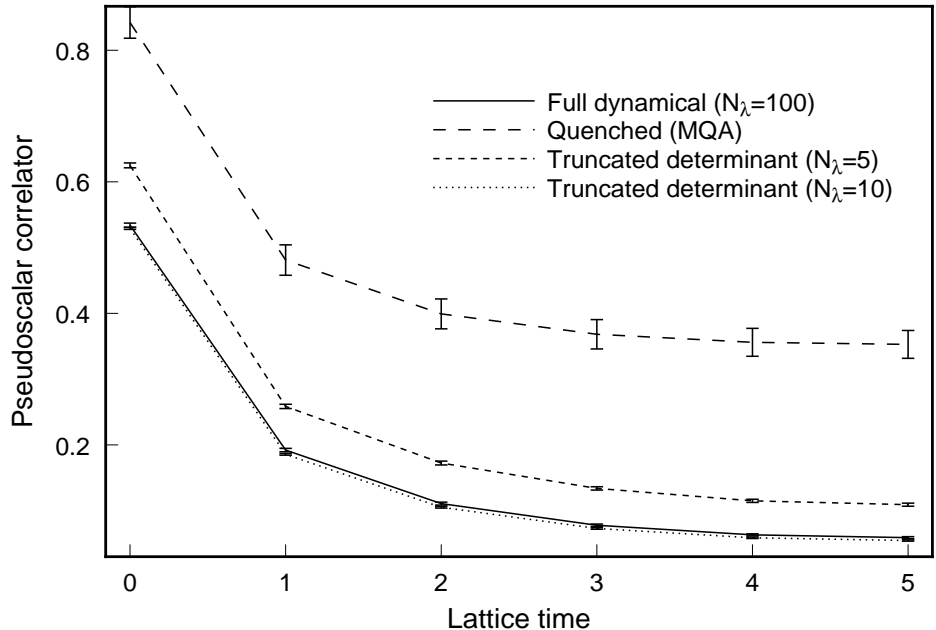,
width=0.95\hsize}
\caption{Comparison of quenched, full dynamical and truncated determinant simulations:
10x10 lattice, $\beta$=4.5, m=0.05}
\label{fig:pionall}
\end{figure}
As we are working at a very small value of quark mass, it is also of interest to study
the effect of the truncated determinant factor on the topological charge distribution
and the string tension 
of the theory. For the quenched simulations on a 10x10 lattice at $\beta=$4.5, the
topological charge, defined as
\begin{equation}
\label{eq:q1def}
Q_{\rm top} \equiv\frac{1}{2\pi}\sum_{P}\sin{(\theta_{P})}
\end{equation}
(where $\theta_{P}$ is the plaquette angle for plaquette $P$), is found to be
concentrated at {\em roughly} integer values, with charges 0 and 1 dominating. The
histogram of topological charge values obtained from 800 quenched 
configurations is shown in Fig. 4. As low  eigenvalues are introduced via
the truncated determinant, the nonzero topological charge configurations are suppressed.
Again, with $N_{\lambda}$=5, the resulting distribution is hardly distinguishable
from the full dynamical result.

\begin{figure}
\psfig{figure=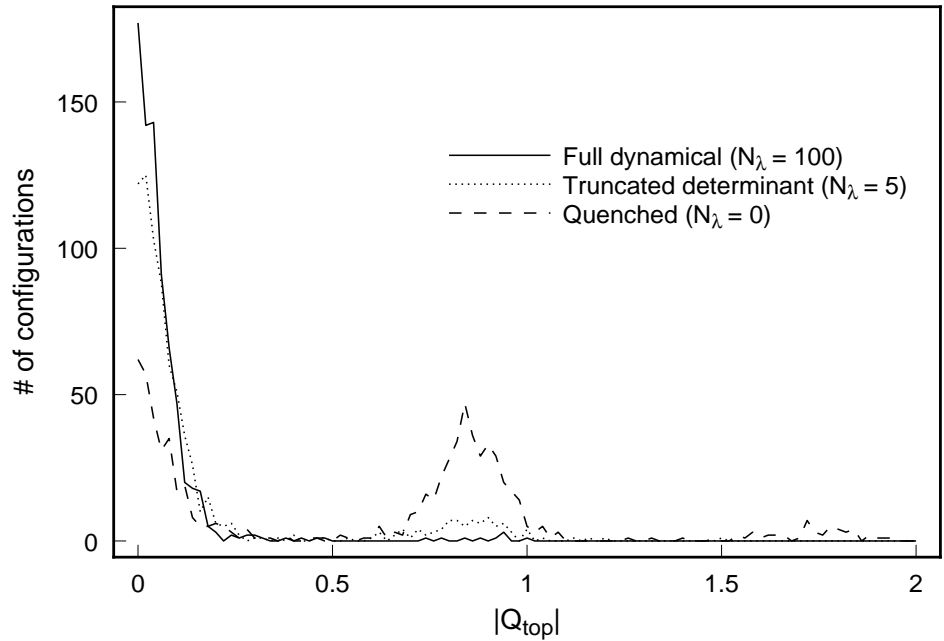,
width=0.95\hsize}
\caption{Topological charge distribution for quenched, 
full dynamical and truncated determinant
simulations: 10x10 lattice, $\beta$=4.5, m=0.05}
\label{fig:topcharge}
\end{figure}

The quark-antiquark potential determined for two different sea
quark masses (bare mass 0.06 and 0.10)
is shown in Fig 5. The calculation was done in the truncated theory on a 16x16 lattice
at $\beta$=4.5 using $N_{\lambda}$=10 eigenvalues. Also shown is the exact result for
the quenched theory and exact update full dynamical theory (at sea quark mass 0.06). 
The small eigenvalues properly represent the large loops induced
by the determinant, leading to a breaking of the string at longer distances for light
quarks.

\begin{figure}
\psfig{figure=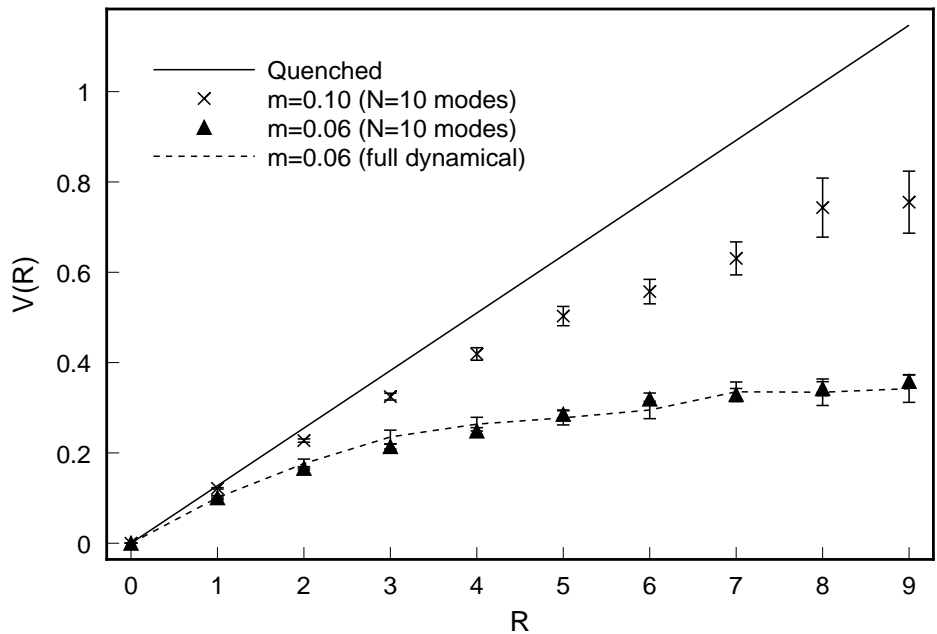,
width=0.95\hsize}
\caption{Quark-antiquark potential in QED2.}
\label{fig:qedstring}
\end{figure}

\section{Calculating Truncated Determinants in Large Sparse Systems}

There are a number of techniques available for the extraction of a limited
number of low eigenvalues of a large sparse linear system: most popular are 
conjugate gradient approaches \cite{conjgrad}, or the Lanczos
technique \cite{golubbook}, suitably modified to guard for the appearance of spurious
eigenvalues \cite{cullwill}. The latter approach has been studied extensively by
Kalkreuter \cite{Kalk}, and has proven to be most well suited for the task at
hand, namely, the accurate extraction of a complete set of low-lying eigenvalues
of $\gamma_{5}(D\!\!\!/-m)$ up to an energy scale which ensures that all the 
important soft chiral dynamics of full QCD is included in the Monte Carlo 
simulation. Typically, on lattices of physically interesting volume in 4D QCD, this 
requires the determination of something on the order of 100 eigenvalues. One
advantage of the Lanczos approach is that it can be pushed through to the 
determination of as many eigenvalues as desired, with a numerical effort which grows
empirically as roughly the square  of the desired energy cutoff (in the portion
of the spectrum corresponding to the physical branch). In particular,
on small lattices, it is relatively straightforward to determine the entire
spectrum, which is useful both for diagnostic purposes and in studying the
systematic effects of an algorithm based on a truncated determinant.

The Lanczos technique is a standard part of the literature in Numerical Analysis
(see, for example, \cite{golubbook}) so we shall give only a very brief review of the
procedure here. Given a hermitian matrix $H$ (in our case, this is just the
matrix $\gamma_{5}(D\!\!\!/-m)$ discussed previously), a series of orthonormal
vectors $v_1, v_2, v_3...$ are generated from a starting vector $w_0\equiv v_1$ by the 
following recursion:
\begin{eqnarray}
  v_{n+1} &=& w_{n}/\beta_{n}  \\
  n &\rightarrow& n+1 \\
  \alpha_{n} &=& (v_{n},Hv_{n})  \\
  w_{n} &=& (H-\alpha_{n}I)v_{n} - \beta_{n-1}v_{n-1} \\
  \beta_{n} &=& \sqrt{(w_{n},w_{n})}
\end{eqnarray}
where $\alpha_{n},\beta_{n}$ are real numbers (by virtue of the hermiticity of $H$),
and the initial conditions are $\beta_{0}=1, v_{0}= 0$. It is straightforward to
verify that the matrix of $H$, in the basis spanned by the Lanczos vectors $v_{n}$
is tridiagonal, with the numbers $\alpha_1,\alpha_2,..$ down the main diagonal,
and the numbers $\beta_1,\beta_2,..$ on the first super- (and sub-)diagonals. If
the Lanczos recursion is carried to order $N$, one is thus led to a $N$x$N$
truncation of the original linear system, and the eigenvalues of the resulting
tridiagonal matrix $T^{(N)}$ represent increasingly accurate approximants to the
eigenvalues of the full system, provided only that the starting vector $w_0$ is
not entirely contained in an invariant subspace of $H$. Typically, $w_0$ is chosen
randomly and one expects that such a random vector will overlap nontrivially with
all the eigenvectors of $H$. 

There are several features of the Lanczos procedure
which appear at first sight problematic but 
which nevertheless turn out not to compromise its efficacy in the present application.
First, degenerate eigenvalues of the original matrix $H$ are not properly handled
(although methods have been devised for circumventing this drawback \cite{golubbook}). 
This turns out to be irrelevant in our QCD application as the generic spectrum
of $H\equiv\gamma_{5}(D\!\!\!\!/-m)$ for a typical gauge configuration encountered in the
course of a Monte Carlo simulation is entirely nondegenerate. Secondly, the
effects of roundoff error in the algorithm can be quite severe, and lead to the
appearance of spurious eigenvalues and to the false duplication of real eigenvalues.
Fortunately a simple and effective cure for this problem, first suggested by 
Cullum and Willoughby \cite{cullwill}, proves to be practical in the gauge theory case
\cite{Kalk}. However, it remains an unfortunate feature of the algorithm that the
number of accurate eigenvalues extracted at level $N$ of the recursion is typically
considerably smaller than $N$. For example, on a 12$^3$x24 lattice at $\beta$=5.9,
the extraction of the lowest 100 eigenvalues of the Dirac operator typically requires
on the order of 10,000 Lanczos sweeps. (The computational cost of a single 
Lanczos sweep is essentially that of the single $D\!\!\!/$ multiplication 
incurred in producing the next Lanczos vector $w_{n} = Hv_{n}+..$.) 
Finally, although the diagonalization of
a tridiagonal matrix is conceptually trivial and efficiently implementable by scalar
algorithms (e.g. by a QL implicit shift algorithm \cite{numrec}), the parallel
implementation of this procedure is not entirely trivial. In our
 simulations, this is essential to avoid a serious bottleneck in the
simulation when tridiagonal matrices of order up to several tens of thousands must
be efficiently processed. We describe below an elegant 
parallel approach to the extraction of the spectrum of $T^{(N)}$. 

In the case of the Dirac operator in QCD, the choice of the random vector $w_0$
used to start the recursion appears to be fairly innocuous (a local source seems
perfectly adequate, for example) . An important check that the spurious eigenvalues
are correctly identified and that the remaining ``good" eigenvalues are sufficiently
converged relies on the gauge-invariance of the individual eigenvalues of
$H$ which can be verified explicitly by recomputing the eigenvalues with
varying degrees of gauge-fixing. We have performed extensive checks to
ensure that the eigenvalue spectrum, and a-fortiori the truncated 
determinant ${\cal D}(N_{\lambda})$, is invariant (typically to at least 
8 significant figures) under gauge transformations
of the input configuration.

The procedure we use to isolate converged eigenvalues of $H$ involves two stages.
First, spurious eigenvalues are identified by the procedure of
Cullum and Willoughby- namely, eigenvalues of the tridiagonal matrix 
$T^{(N)}$ are compared with those of the matrix $\hat{T^{(N)}}$ obtained by
deleting the first row and column of  $T^{(N)}$ and removing all common 
eigenvalues of the two systems. Secondly, converged eigenvalues are
identified by requiring either duplication of the remaining good eigenvalues
or stability within a preassigned precision level when eigenvalues are
compared at recursion level $N-N_{\rm gap}$ and $N$. Typically we insist
on a precision level of at least 10$^{-5}$ and choose $N_{\rm gap}$=100.
The above procedure requires the resolution of the central part of the eigenvalue
spectrum for four large tridiagonal matrices (of dimension $N$,$N$-1,$N-N_{\rm gap}$,
and $N-N_{\rm gap}$-1, respectively).  It is therefore highly desirable to
perform these diagonalizations in a way that allows parallelization. 

  The key to extracting the central part of the spectrum of a tridiagonal matrix in
a parallel machine lies in the Sturm Sequence property of such matrices \cite{golubbook},
valid provided none of the subdiagonal entries $\beta_{n}$ vanish, as is 
certainly the case for generic gauge configurations. Let $p_{n}(\lambda)$ be
the secular determinant ${\rm det}(T^{(N)}_{n}-\lambda I)$ of the $n$x$n$ principal
submatrix $ T^{(N)}_{n}$ of $T^{(N)}$.  Then the number of eigenvalues of
$T^{(N)}$ less than any preassigned $\lambda$ equals the number of sign
changes in the sequence $p_{0}(\lambda),p_{1}(\lambda),...p_{N}(\lambda)$.
As the $p_{n}(\lambda)$ can readily be calculated by a simple two term
recursion relation, an obvious bisection procedure can be used to determine
the $n$'th eigenvalue of $T^{(N)}$ at any desired level of precision. Moreover,
the task of extracting different eigenvalues can be assigned completely
independently to separate processors, provided global access to the elements
$(\alpha_{i},\beta_{i})$ of $T^{(N)}$ is arranged.

\begin{figure}
\psfig{figure=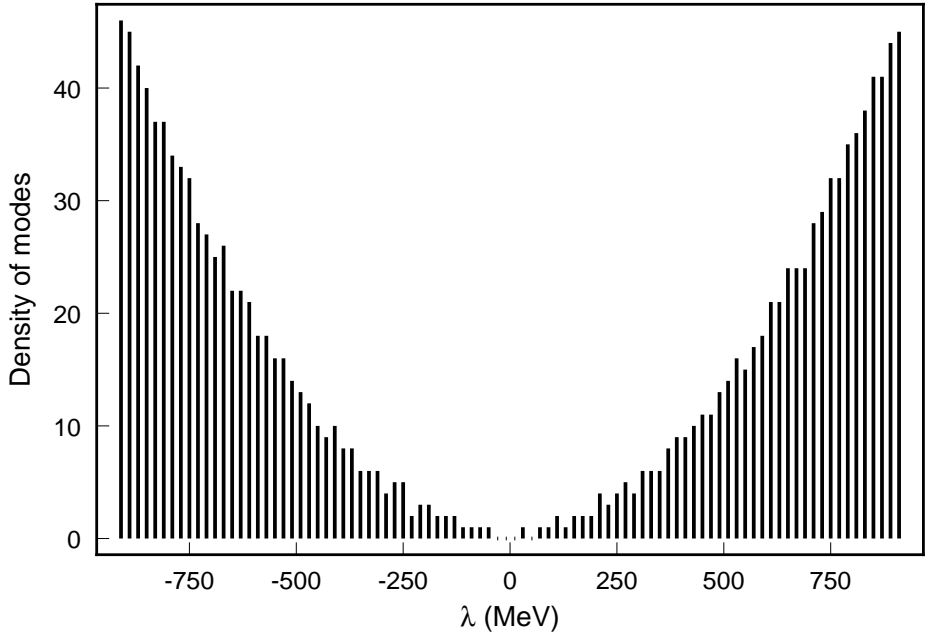,
width=0.95\hsize}
\caption{Spectral density,$\beta$=5.9,$\kappa$=0.1587}
\label{fig:spec12x24}
\end{figure}

In Fig 6 we show the spectrum of $H$ for a typical configuration on a 12$^3$x24 lattice at
$\beta$=5.9, $\kappa$=0.1587.  In this case an extended  Lanczos recursion was carried out
up to order $N$=78000, yielding 1478 converged eigenvalues in the central region 
of the spectrum. Converting the gauge-invariant  eigenvalues of $H$ to a physical
energy scale (using the scale $a^{-1}$=1.78 GeV from the
charmonium spectrum), this corresponds to all quark eigenmodes up to 970 MeV. The
energy reach as a function of number of Lanczos sweeps for this lattice is
shown in Fig 7. In the 
simulations reported below, we have typically used 9500-12000 Lanczos sweeps
(with slightly different tunings of the Cullum-Willoughby procedure) and included
the lowest 100 (i.e. 50 positive and 50 negative) eigenvalues of $H$ in the update
procedure. This cutoff corresponds to inclusion of quark eigenmodes up to an
energy of approximately 370 MeV.  Lanczos recursion to order $N\sim$10000 requires
(for a 12$^3$x24 lattice) 
about an hour on 64 nodes of the Fermilab ACPMAPS system.
\begin{figure}
\psfig{figure=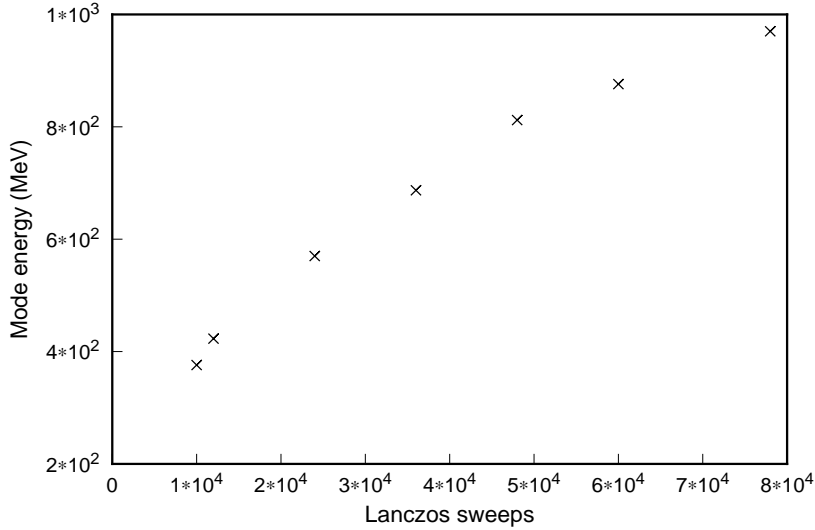,
width=0.80\hsize}
\caption{Lanczos convergence,$\beta$=5.9,$\kappa$=0.1587}
\label{fig:lancconv}
\end{figure}

\begin{figure}
\psfig{figure=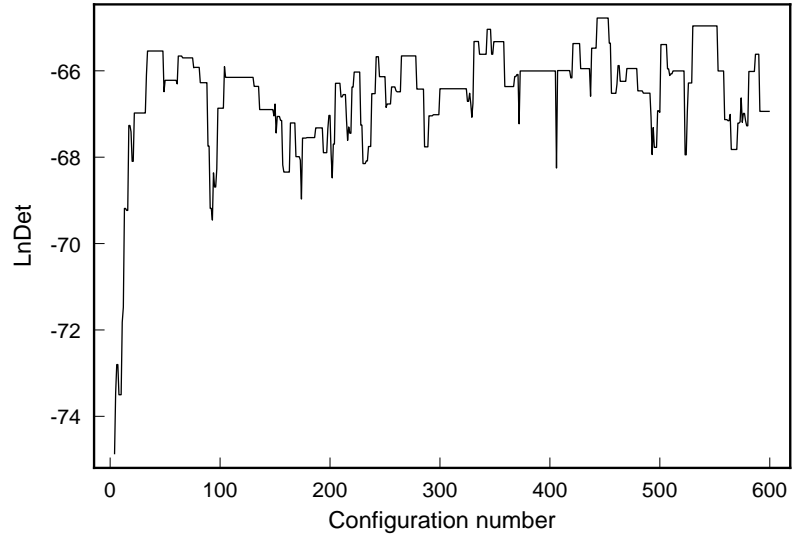,
width=0.80\hsize}
\caption{Determinant relaxation,$\beta$=5.7,$\kappa$=0.1685,$N_{\lambda}$=30}
\label{fig:detrelax_k}
\end{figure}

\begin{figure}
\psfig{figure=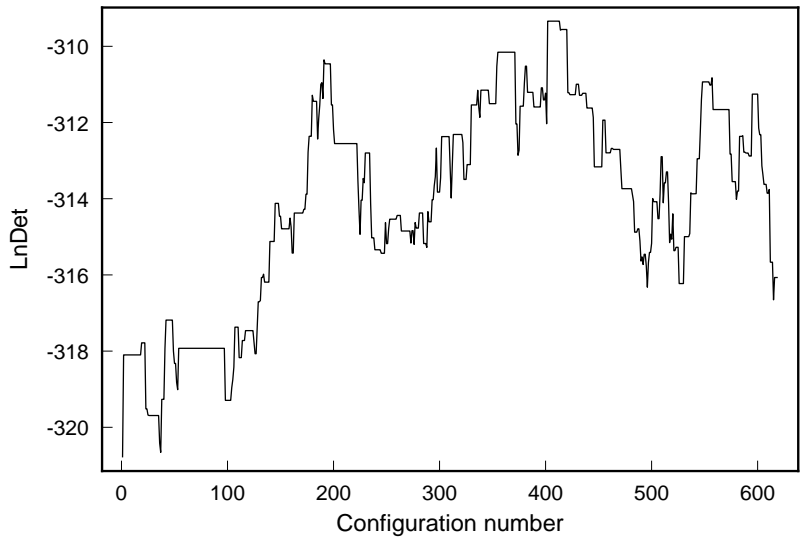,
width=0.80\hsize}
\caption{Determinant relaxation,$\beta$=5.9,$\kappa$=0.1587,$N_{\lambda}$=100}
\label{fig:detrelax_b}
\end{figure}

The relaxation of the determinant from its typical quenched value to the equilibrium
value appropriate for the system simulated with the truncated determinant is shown
in Figs. 8 and 9. On a small lattice, (6$^4$ at $\beta=5.7$, $\kappa$=0.1685) 
a dynamical run including $N_{\lambda}=$30 eigenvalues 
(or up to about 500 MeV in quark mode energy) was performed,
with a determinant update accept/reject every 3 heat-bath sweeps of the lattice. The
resulting evolution of ${\cal D}(N_{\lambda})$ starting from a quenched configuration is
shown in Fig 8. The same evolution is shown for a run on a 12$^3$x24 lattice at
$\beta$=5.9, $\kappa$=0.1587 with $N_{\lambda}=$100 in Fig 9, 
and with determinant accept/reject
every 2 heat-bath sweeps.

 The procedure we have used does not show a very strong dependence of the acceptance rate
on the quark mass (fortunately!). For the heaviest quark mass studied at 
$\beta=$5.9, $\kappa=$0.1570 (slightly lighter than the strange quark), the
 acceptance rate for a run with determinant accept/reject performed every two gauge
 heat bath sweeps was 40\%. For the lightest two quark masses ($\kappa$=0.1587 and
0.1597) two separate runs were performed with determinant accept/reject steps separated
by either one or two heat-bath sweeps. For the heavier case, $\kappa$=0.1587 
(corresponding to a pion mass  of around 400 MeV), the acceptance was on the
order of 37\% for new configurations separated by 2 heat-bath sweeps, and 57\% for 
configurations separated by a single heat-bath sweep. For the lighter mass, $\kappa$=0.1597,
(corresponding to a pion mass of around 280 MeV)
the acceptance was 30\% for new configurations separated by 2 sweeps, and 43\% for 
those separated by a single heat-bath sweep.

\section{Simulations in QCD4}

To test the efficacy of the truncated determinant approach to unquenched QCD
we have performed some preliminary runs on a 12x$^3$x24 lattice at $\beta$=5.9,
for two degenerate flavors of dynamical quarks at hopping parameter values $\kappa=$ 0.1570,  0.1587 and 0.1597. For the heaviest mass,
$\kappa=$0.1570, propagators were computed for every fifth configuration (60 in all),
and determinant
accept/reject performed after every two heat-bath gauge updates. 
For the lighter two masses, we also performed parallel runs with determinant accept/reject
after every heat-bath sweep and with propagators measured every tenth configuration.
 Our results are based on 104 propagators for $\kappa=$0.1587
and 88 propagators for $\kappa=$0.1597. As fluctuations in the large distance behavior
at light quark masses grow, the run at $\kappa=$0.1597 is continuing and results with
much higher statistics will be presented in a later work.

  Pseudoscalar meson masses were determined by measuring meson correlators with a 
 smeared source and local sink and doing a fully correlated Euclidean time fit in
the time window 8-11. At kappa values $\kappa$=0.1570, 0.1587, and 1597, the pion
masses were found to be (lattice units) $M_{\pi}=$ 0.339$\pm$0.011, 0.241$\pm$0.019 and
 0.198$\pm$ 0.069. 
The extrapolation to zero pion mass
gives a critical kappa value of $\kappa_{c}\approx$0.1602, 
slightly higher than the pure quenched value of $\kappa_{c}$=0.15972 found in
previous calculations \cite{fbpaper}. With this critical kappa, the pion
 mass corresponding to our lightest case should be about 0.157 (lattice units),
 which converts to about 280 MeV using the scale determined for the quenched theory
at this beta. We do not expect the scale in the truncated determinant simulation
to be much changed from the quenched theory as the shift in beta in full dynamical
simulations (see \cite{deGrand})  is due to a large
 logarithm accumulated from quark modes all the way up to the lattice cutoff,
 whereas the truncated determinant included here only takes into account the infrared modes
up to 370 MeV.  
  Higher statistics are presently being accumulated
at the lightest quark mass, where the fluctuations in the large time meson correlators
are largest. 

An important aspect of the low energy chiral dynamics of QCD is the response
of the topological structure of the theory to the presence of light dynamical 
quarks. Integrating the vacuum expectation value of the U(1) axial anomaly in 
QCD yields immediately, in the continuum, the anomalous chiral Ward identity
\begin{equation}
 m_{q}\int d^{4}x <\bar{\psi}(x)\gamma_{5}\psi(x)> = \frac{1}{32\pi^{2}}\int d^{4}x F_{\mu\nu}\tilde{F}_{\mu\nu} \equiv Q_{\rm top}
\end{equation}
We may therefore define a topological charge on the lattice by simply evaluating the
left-hand-side of the above equation configuration by configuration. The advantage
of this definition is that the required information is already immediately accessible:
the Euclidean vacuum expectation value  $\int d^{4}x<\bar{\psi}(x)\gamma_{5}\psi(x)>$ reduces to
the trace of the inverse of $H\equiv\gamma_{5}(D\!\!\!/-m)$, the low eigenvalues of which
were extracted in the course of the simulation. Namely, we have the following lattice
definition of $Q_{\rm top}$
\begin{equation}
 Q_{\rm top} \equiv \frac{1}{2\kappa}(1-\frac{\kappa}{\kappa_c})\sum_{i=1}^{N}\frac{1}{\lambda_{i}}
\end{equation}
where the matrix dimension of $H$ is $N$ and $\lambda_{i}$ are the eigenvalues of $H$.

\begin{figure}
\psfig{figure=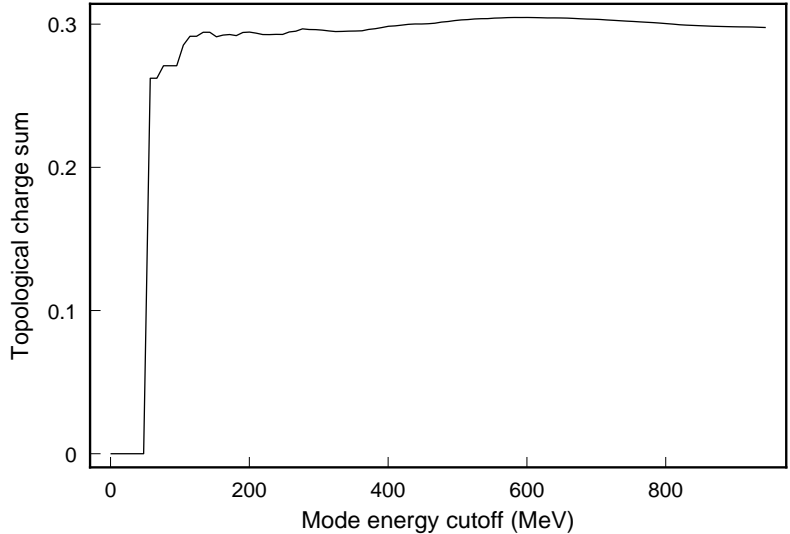,
width=0.80\hsize}
\caption{Convergence of the topological charge spectral sum for a typical configuration}
\label{fig:topconv}
\end{figure}

As the larger eigenvalues occur roughly as equal and opposite pairs, this sum actually
saturates quickly at the low end, and the 100 eigenvalues computed already in the
course of the simulation suffices to determine $Q_{\rm top}$ to a few percent. An
example of the convergence  of this spectral sum, with the mode eigenvalues converted to
a physical energy scale (recall that the inclusion of 100 eigenvalues corresponds to
modes up to about 370 MeV), is shown in Fig \ref{fig:topconv} for a typical
configuration.

\begin{figure}
\psfig{figure=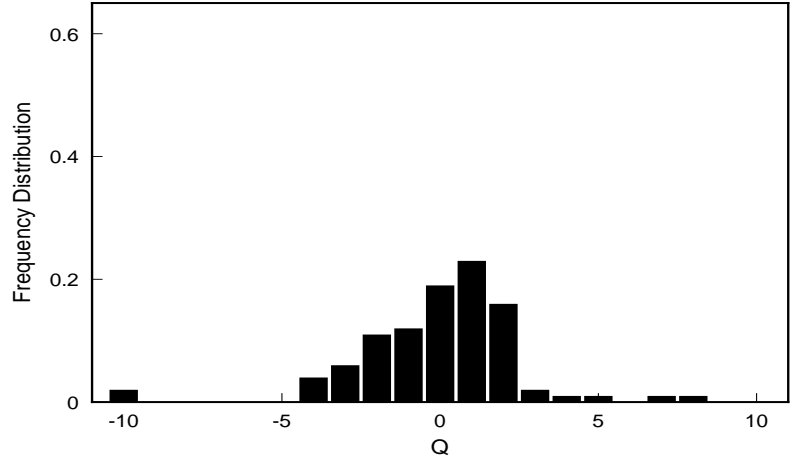,height=2.5in,width=0.80\hsize}
\caption{The topological charge frequency distribution computed 
for $\kappa$ = .1587 using 100 decorrelated quenched
QCD configurations on a 12$^3$x24 lattice at $\beta$ = 5.9.}
\label{fig:topk1587qa}
\end{figure}

\begin{figure}
\psfig{figure=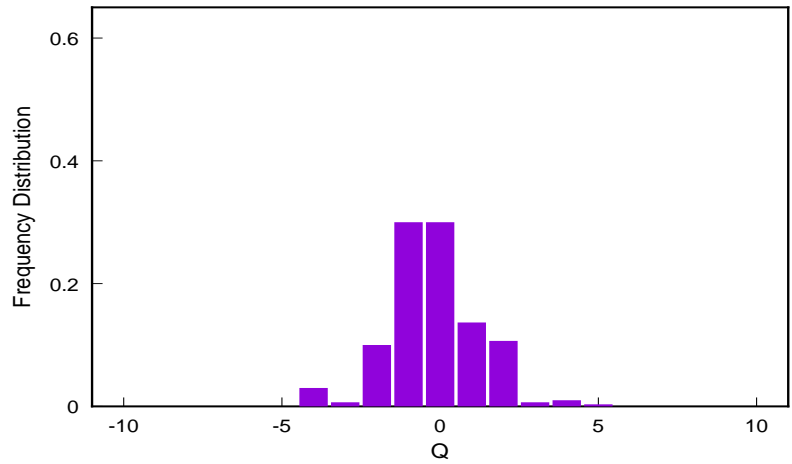,height=2.5in,width=0.80\hsize}
\caption{The same topological charge frequency distribution 
for 300 configurations generated by the truncated determinant algorithm  
with sea quark mass $\kappa =$ .1587
on a 12$^3$x24 lattice at $\beta =$ 5.9.}
\label{fig:topk1587}
\end{figure}

With the definition (18), the qualitative effect of the quark determinant on the 
topological charge distribution can readily be studied. Two effects are clearly
visible in our data:

(1) The real mode artifacts characteristic of quenched Wilson gauge  theory,
and which underly the increasingly frequent appearance of exceptional configurations
as one goes to lighter quark masses, do not occur. 
Configurations with very small eigenvalues of $H$ are suppressed by  the
determinant factor which is driven strongly negative for any such configuration.
In Fig \ref{fig:topk1587} this effect is seen comparing the topological charge
distribution for a dynamical run at $\kappa=0.1587$ (including the lowest 100 modes)
with a completely quenched run at the same valence quark mass, Fig \ref{fig:topk1587qa}.
The distributions
are broadly similar (although the dynamical one is slightly narrower) but the 
outlying points corresponding to very large values of $Q_{\rm top}$ (i.e to
the appearance of an exactly real mode of the Wilson-Dirac 
operator  very close to the chosen kappa value) are eliminated in the dynamical run.
In fact in the quenched case 
there are several outlying points (the furthest out at $Q_{\rm top}$ = -91.6)
not shown on the figure. For the dynamical run only charges $|Q_{\rm top}|< 5$ are seen.


\begin{figure}
\psfig{figure=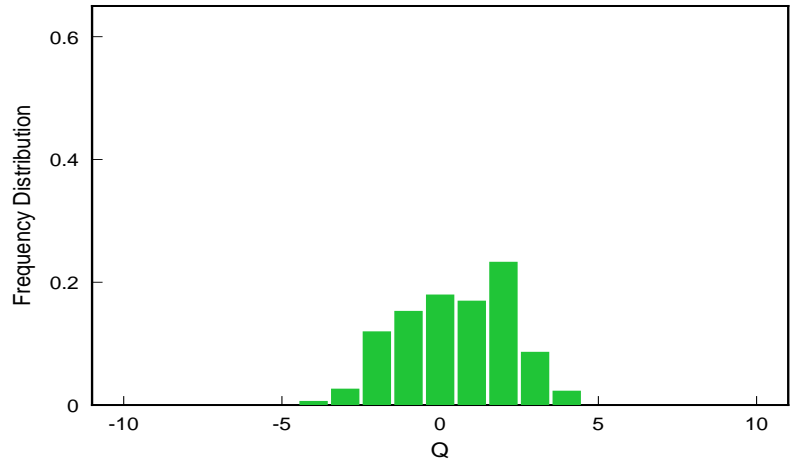,height=2.5in,width=0.80\hsize}
\caption{The topological charge frequency distribution 
with sea quark mass $\kappa =$ .1570. 
(Other parameters as in the previous figure.)}
\label{fig:topk1570}
\end{figure}
\begin{figure}
\psfig{figure=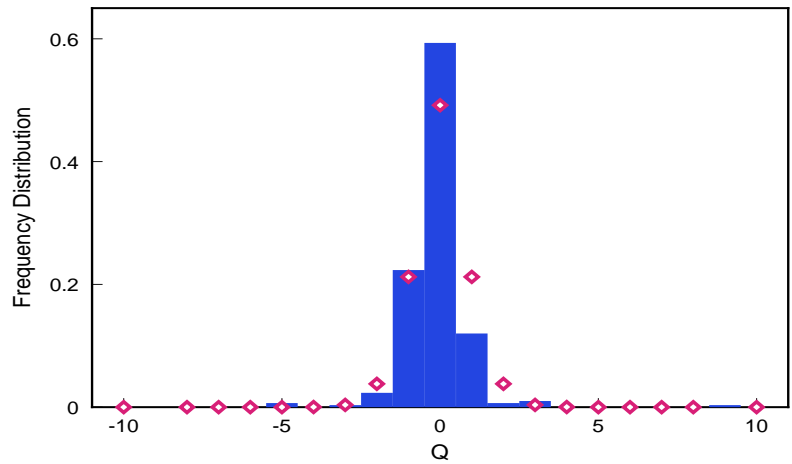,height=2.5in,width=0.80\hsize}
\caption{The same topological charge frequency distribution 
with sea quark mass $\kappa =$ .1597.
The expected distribution from chiral
perturbation theory is shown with diamonds.}
\label{fig:topk1597}
\end{figure}

(2) Nonzero topological charge must be suppressed in the chiral limit of 
vanishing quark mass, so we expect that the histogram of measured topological charges will
narrow as one approaches $\kappa_c$. 
This effect is shown in Figs. \ref{fig:topk1570} and \ref{fig:topk1597}
where the topological charge distribution is compared for two dynamical runs at the
lowest and highest quark masses studied ($\kappa$= 0.1597 and 0.1570). The
narrowing of the distribution for the lighter mass is immediately apparent. For comparison 
we plot the analytic result predicted by the chiral analysis of Leutwyler and Smilga \cite{LeutSmil}
in the light quark limit for two degenerate flavors:
\begin{equation}
 P(Q)= I_{Q}(x)^{2}-I_{Q+1}(x)I_{Q-1}(x)
\end{equation}
with $x\equiv \frac{1}{2}VF_{\pi}^{2}M_{\pi}^{2}$. Here $V,F_{\pi},M_{\pi}$ denote the
lattice space-time volume, the pseudoscalar decay constant and the pion mass respectively,
all in lattice units. For $\kappa$=0.1597 we have taken $M_{\pi}$=0.15 and $F_{\pi}$=0.07
(the latter number is extrapolated from high statistics quenched runs for this lattice
\cite{fbpaper}).

\begin{figure}
\psfig{figure=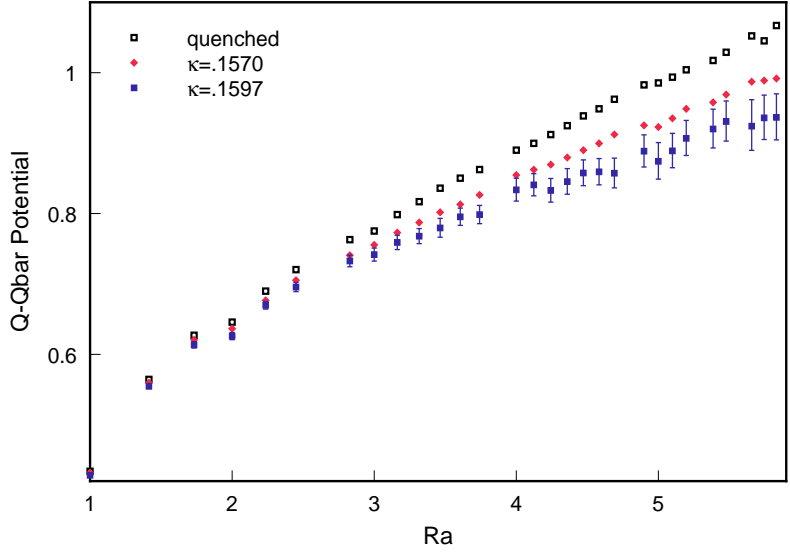,height=3.0in,width=0.80\hsize}
\caption{Quark-antiquark potential for quenched 
theory and dynamical theory with sea quark kappa values 0.1570, 0.1597. 
Error bars shown only for lightest mass  
(errors are largest for this case).}
\label{fig:qqbarqcd}
\end{figure}

The presence of low-momentum virtual sea-quark modes in the simulation should
result in screening of the quark-antiquark potential 
extracted from Wilson loops
at large distance. In Fig \ref{fig:qqbarqcd} the potential obtained in the quenched 
theory on a 12$^3$x24 lattice at $\beta=$5.9 (200 configurations) is compared with 
that calculated from our dynamical configurations at the lightest sea quark mass
($\kappa=$0.1597). The effect of screening is clear although asymptotic flattening of the 
potential on this lattice occurs at distances  where statistical fluctuations as well
as finite volume effects dominate. 

\section{Matching the High Eigenvalues}

Because QCD in four dimensions is only renormalizable, not 
super-renormalizable, the fluctuations of 
the fermion determinant are significant at all physical scales.  
Therefore, unlike QED in two dimensions, 
it can not in general be sufficient to  compute accurately only the low
eigenvalues of the fermionic determinant.
Fortunately the short distance behaviour of QCD is very well understood.
This should allow the identification of the important degrees of freedom
and lead to a method for including the effects of the higher eigenvalues
into the Monte Carlo process.  
In particular, we know that for sufficiently high momentum scales 
this physics should be accurately described by an improved gauge action
 involving Wilson loops on short distance scales only.

The fermion determinant can be separated into two pieces
\begin{equation}
\ln {\rm det} H = [{\rm Tr} \ln H]_{low \;\lambda} + [{\rm Tr} \ln H]_{high \;\lambda} 
\end{equation}
where the lowest $n_{\rm cut}$ eigenvalues are directly calculated 
and included in the Monte Carlo updating procedure. The contribution
of the vast majority of the larger eigenvalues can be included by
some approximation to the high end that 
(1) matches onto the low eigenvalue results without gaps or 
double counting, (2) is controlled  and (3) becomes exact in the
continuum limit. We can define the difference $\Delta$ between the approximate 
action, denoted $S_a$, and 
the exact contribution of the high eigenvalues,
$S_t \equiv [{\rm Tr} \ln H]_{high \;\lambda} $ 
as follows:  
\begin{equation}
 \Delta = S_t - S_a
\end{equation}
Any acceptable method must ensure that this difference is small 
($\le 1$) for each configuration. 
Therefore, we will demand that the variance of $\Delta$ for any set 
configurations is less than unity. Actually, we will find that a
relatively simple effective loop action yields values of $\Delta$
considerably less than unity for interesting values of $n_{\rm cut}$.

Two numerical methods suggest themselves for  calculating the 
high eigenvalues of the fermion determinant:
\begin{itemize}
\item The multiboson approach of L\"{u}scher\cite{multiboson}.
\item Using a small number of gauge loops to model the determinant as proposed
by  Sexton and Weingarten \cite{Wein97}, and Irving and Sexton \cite{Irving97}.
\end{itemize}

One method to compute the high eigenvalues which is guaranteed to succeed
is the multiboson approach of L\"{u}scher\cite{multiboson}. 
Define 
\begin{equation}
P_{eff}(U) \equiv [{\rm det(D+m)}]^{n_f} \exp{(-S_g(U))} 
\end{equation}
and 
\begin{equation}
H = \gamma_5 {\rm(D+m)}/[{\rm c_m(8+m)}]\;\;\; (c_m \geq 1) 
\end{equation}
where $c_m$ is chosen so that the eigenvalues of $H$ are in the interval $(-1,1)$.
Consider two flavors of light Wilson-Dirac quarks ($n_f = 2$).
L\"{u}scher chooses a sequence of polynomials $P_{n}(s)$ of even degree  n such that
\begin{equation}
\lim_{n\rightarrow \infty} P_{n}(s) = 1/s \;\;\;{\rm for \;\;all\;\;} 0 < s \leq 1
\end{equation}
then
\begin{equation}
 {\rm det} H^2 = \lim_{n\rightarrow \infty}[ {\rm det} P_{n}(H^2)]^{-1}
\end{equation}
Choose polynomials such that complex roots $z_1 \dots z_n$ come in complex
conjugate pairs (non real) so that  $\sqrt{z} = \mu + i \nu$.
Then 
\begin{equation}
P(H^2) = const \prod_{k=1}^{n} [(H-\mu_k)^2+\nu_k^2]
\end{equation}
and 
\begin{equation}
 {\rm det} H^2 = \lim_{n\rightarrow \infty}\prod_{k=1}^n {\rm det}[(H-\mu_k)^2+\nu_k^2]^{-1}
\end{equation}
Hence we can finally write 
\begin{equation}
 P_{eff}(U) = \lim_{n\rightarrow \infty}\frac{1}{Z_b} \int D\phi D\phi^{\dagger}
 \exp{-S_b(U,\phi)}
\end{equation}
where the bosonic action is given by 
\begin{equation}
S_b = S_g(U) +  \sum_{k=1}^n \sum_{x}|(H-\mu_k)\phi_k(x)|^2 + \nu^2|\phi_k(x)|^2 .
\end{equation}
To estimate how many boson fields are required to represent the original action
to a fixed accuracy in the range ($\epsilon < s \leq 1$) L\"{u}scher 
considered polynomials of the Chebyshev type (denoted T).
Defining $u=(s-\epsilon)/(1-\epsilon)$ and $\cos \theta = 2u-1$, 
$T^{\star}_r(u) = \cos(ru)$ and we can write
\begin{eqnarray}
P(s) &=& [1+\rho T^{\star}_{n=1}(u)]/s \\ 
R(s) &=& [P(s) - (1/s) ]s (\epsilon < s \leq 1) \nonumber
\end{eqnarray}
where $\rho$ is chosen so that the $P(s)$ is finite as $s\rightarrow 0$.
The error is given by:
\begin{equation}
|R(s)| \leq 2 (\frac{1-\sqrt{\epsilon}}{1+\sqrt{\epsilon}})^{n+1} 
\end{equation}
where n is the number of boson fields and the fit is cutoff 
for eigenvalues below $\epsilon$. 
Therefore, the convergence is exponential with rate $2\sqrt{\epsilon}$
as $n\rightarrow \infty$. 

The main practical problem with this 
multiboson method is that it requires an increasingly large number of boson 
fields as the quark mass becomes lighter. As $m_q \rightarrow 0$, we must 
take $\epsilon \rightarrow 0$, but to obtain a fixed level of accuracy 
we must hold $2\sqrt\epsilon n$ fixed and hence $n$ increases without bound.

However the multiboson method matches nicely onto the calculation of 
low eigenvalues discussed  previously. 
In our truncated determinant method, the cutoff $\epsilon$ for the
multiboson method is set by the highest eigenvalue of $s = (\gamma_5 (D+m))^2$
which is explicitly included in the low end calculations.  Hence 
it does not explode as the quark mass goes to zero. The combination of 
methods remain accurate for all quark masses. For example, for $\beta$ = 5.9
on a 12$^3$x24 lattice with direct inclusion of the lowest 100 eigenvalues,
the associated cutoff for the multiboson simulation of the high eigenvalues
is $\sqrt{\epsilon} \approx$ 0.035 independent of the light quark mass.

Furthermore, the error associated
with the inaccurate behaviour of the polynomial fit in the 
range $0 < s < \epsilon$ can be corrected as low eigenvalues are computed
for every configuration update.  We obtain a reweighting term,
\begin{equation}
 \Delta S_b =  \sum_{i=1}^{n_{\rm cut}}\ln {(\lambda_i^2  P(\lambda_i^2))}
\end{equation}
which can be included to eliminate errors in the region $0 < s \leq \epsilon$.

Using the multiboson method for the high end of the determinant satisfies 
all our requirements and completes the algorithm. 
However it is interesting to study if we can reduce the total required 
computations even further using a more physical approach to the high eigenvalues.
First, consider how many of the high eigenvalues we are computing actually
have physical information and are not just lattice artifacts. For example,
for a 12$^3$x24 lattice with $\beta$ = 5.9 and $\kappa =$ .1587 there 
are 497,664 total eigenvalues of the Wilson-Dirac operator.  We can explicitly
calculate the number of eigenvalues less than some high energy cutoff. Using 
$1 \gev$ we have approximately 1500 eigenvalues (0.3\%).  For a
fixed volume V and quark mass $m_q$ only a decreasing fraction of the eigenvalues
are below a fixed physical scale as $\beta \rightarrow 0$. Therefore, most of the range of large s fit in L\"{u}scher's  multiboson method
is  physically unimportant. 

This suggests a more physically motivated method for dealing with the high eigenvalue
part of the fermion determinant, in which one approximates the ultraviolet contribution to
 the quark determinant with an effective gauge action: 
\begin{equation}
[{\rm Tr} \ln H]_{high \;\;\lambda} \approx \sum_{i=0}^{i_{max}} \alpha_i L_i
\end{equation}
where each $L_i$ is a set of gauge links which form a closed path. The 
natural expansion is in the number of links. 
For zero links we have 
$L_0$, which is just a constant, for four links we have a plaquette, and six
links give the three terms found in considerations of improved gauge 
actions \cite{improvact}.

This idea was originally proposed by Sexton and Weingarten\cite{Wein97} and
studied in more detail by Irving and Sexton\cite{Irving97}. These 
studies were  done on a 6$^4$ lattice at $\beta$ = 5.7 with Hybrid Monte-Carlo full QCD 
simulations (with a heavy sea quark). Their results were rather discouraging. 
It was hard to get a good approximation to the determinant with a closed set of loops 
and they needed large loops to even approach a reasonable 
fit\cite{Irving97}.

There are however two important differences between their study and our situation.
\begin{itemize}
\item They simulated the {\em whole determinant}, while here we only need to
approximate the eigenvalues above some cutoff. Hence  
we would expect the small loops to dominate at least for sufficiently high
cutoff.
\item They used an approximate procedure to  estimate stochastically
  the logarithm of the determinant needed,
while we are exactly computing {\em all} eigenvalues for this study.
\end{itemize}
It turns out that these differences are critical, as using 
approximately the same lattices (and with even lighter quarks) we find an excellent 
approximation to the high end with only small loops. 

\begin{figure}
\psfig{figure=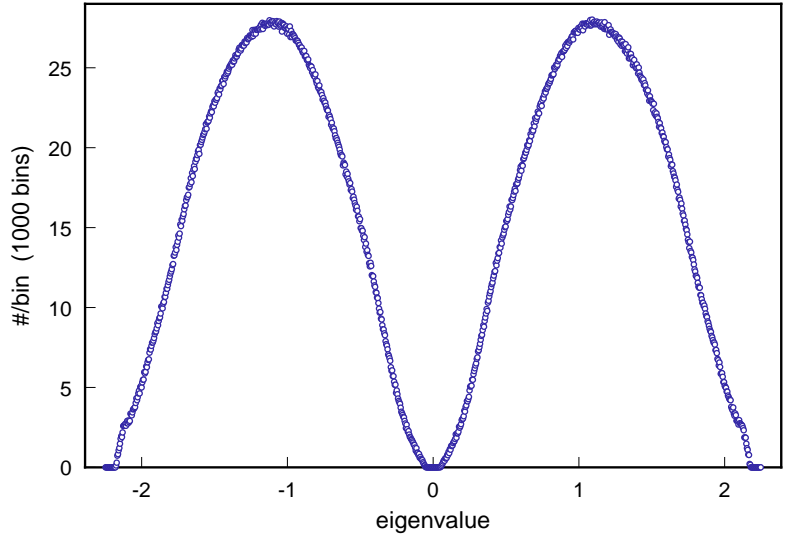,
width=0.80\hsize}
\caption{Spectrum of eigenvalues for $6^4$ lattice,$\beta$=5.7,$\kappa$=0.1685,
$N_{\lambda}$=30}
\label{fig:eigall_box}
\end{figure}

We generated a set of 75 configurations on a 6$^4$ lattice  
at $\beta$ = 5.7 and $\kappa =$ .1685. We included the lowest 30 eigenvalues
(which corresponds to a physical cutoff of approximately $\simeq$350 Mev)
in the Monte Carlo accept/reject step in the generation of these 
independent configurations.
The spectrum of eigenvalues is shown in Fig. \ref{fig:eigall_box}.

\begin{figure}
\psfig{figure=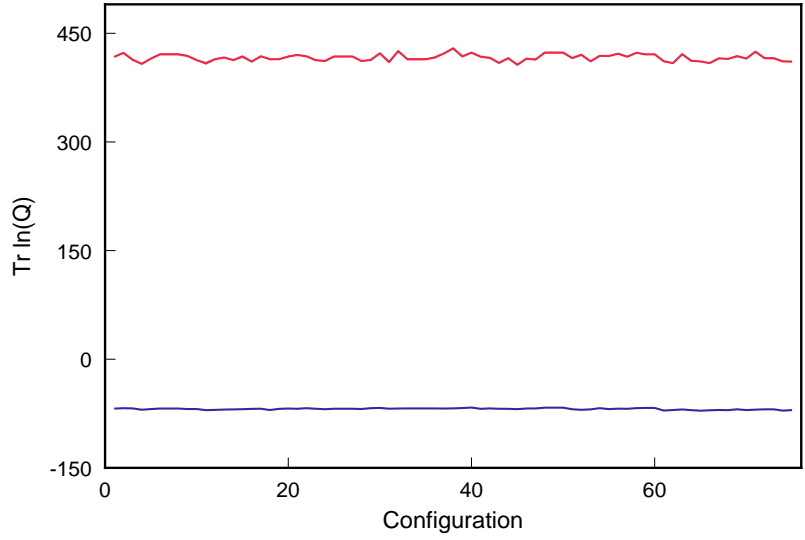,
width=0.80\hsize}
\caption{Fluctuations in the low ($n < 30$) and high eigenvalues for
the same configurations as previous figure.}
\label{fig:hilow_box}
\end{figure}

We can see the importance of the higher eigenvalues by separating the 
high and low part of the fermion determinant for each configuration.
This is shown in Fig. \ref{fig:hilow_box}. Unlike the case of QED2
(cf. Fig 1) it is apparent that the UV contribution to the determinant
definitely involves large fluctuations. Of course, the issue here is
just whether these fluctuations are well described by a simple effective gauge action.

\begin{figure}
\psfig{figure=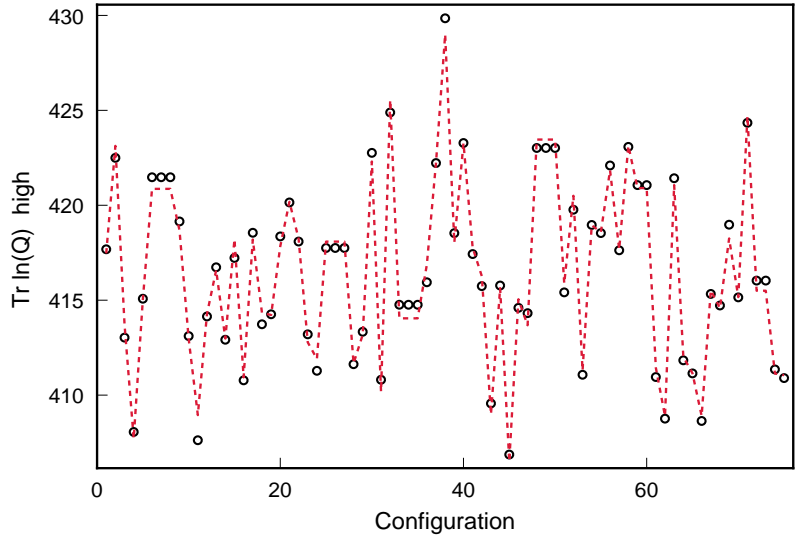,
width=0.80\hsize}
\caption{Fluctuations in the high eigenvalue piece of ${\rm Tr} \ln H$ are
indicated by a dashed line. Fluctuations in
the fitted effective gauge action (6 links and less) 
are indicated with open circles}
\label{fig:det_hilo}
\end{figure}

Considering only the high eigenvalues, an excellent fit to the fluctuations
is obtained including four and six link closed loops. The variance of the
fit is 0.265. The comparison between the fluctuations in the exact and
approximate actions for the high eigenvalue piece is shown in Fig \ref{fig:det_hilo}.

As expected, if only the plaquette term is included the 
variance is larger (2.25) and we must move the low eigenvalue cutoff
to $N=50$ ($\approx$ 700 MeV) to reduce the variance below one.
The results for various cutoffs and terms included are shown in 
Table 1.

\begin{table}
\begin{center}
\begin{tabular}{|c|c|c|c|c|}
\hline
\multicolumn{1}{|c|}{$n_{\lambda}$ cut}
&\multicolumn{1}{c|}{$\lambda$ (MeV)}
&\multicolumn{1}{c|}{4 link fit}
&\multicolumn{1}{c|}{6 link fit}
&\multicolumn{1}{c|}{6 link fit} \\
& & & \multicolumn{1}{c|}{w\/o WL} & \multicolumn{1}{c|}{with WL} \\ 
\hline
0 & 0 & 4.98 & 1.074 & 0.835 \\
$\pm 15$ & 340 & 2.25 & 0.2652 & 0.233 \\
$\pm 50$ & 700 & 0.940 & 0.0564 & 0.0491 \\
$\pm 250$ & 1,210 & 0.0733 & 0.0695 & 0.0641 \\
$\pm 1250$ & 2,220 & 0.138 & 0.0198 & 0.0180 \\
\hline
\end{tabular}
\label{tbl:chisq}
\caption{Fits to the high eigenvalues of the quark determinant by various
small gauge loops (WL denotes a Wilson line passing through the entire lattice).}
\end{center}
\end{table}

Although more study is required this second method looks very attractive
for dealing with the high end of the fermion determinant in full 
QCD with light dynamical quarks. Simulations would be performed by
 including the predetermined effective gauge action $S_a$ in the
 gauge updates and computing the infrared part of the determinant as
 in the truncated determinant simulations described in this paper.

In summary, we have found that there are at least 
two viable methods to deal with the 
contribution of the fermion determinant not computed explicitly in
a truncated determinant approach. 

\section{Conclusions}

We have proposed an algorithm for Monte Carlo simulation of
full QCD with light dynamical quarks which has as its central
feature a separation of the quark determinant into products
over low and high eigenvalues. This separation is a direct reflection
of the different physical roles played by these two sectors,
with high eigenvalues (small quark loops) having the primary effect of
modifying the gauge interaction strength, while the low eigenvalues (large quark
loops) determine the long-range chiral structure of the theory.
Our procedure for generating full QCD configurations then entails
an exact calculation of some fixed number of the lowest lying
eigenvalues of  $H\equiv \gamma_{5}(D\!\!\!/(A)-m)$, using a Lanczos algorithm. This 
careful treatment of low Dirac eigenmodes is motivated by the 
conviction that the most essential differences between quenched
and full QCD reside in their long-range chiral structure and 
associated topological properties.

The complete formulation of our proposed algorithm also allows various
methods of incorporating the higher eigenvalues which were omitted
in the Lanczos calculation. For example, the algorithm matches cleanly onto
the multiboson approach. We have also explored a particularly promising
approach for incorporating the correct ultraviolet behavior
 by the use of a sum over relatively 
small Wilson loops to represent the high-eigenvalue contribution
to $\ln {\rm Det}\;H$. By doing a complete diagonalization on $6^4$ gauge
lattices, we found that this Wilson loop representation works well
with only a few small Wilson loops (4-link and 6-link) included.
Here, the small loop approximation to the truncated  $\ln {\rm Det}\;H$
only succeeds when the lowest eigenvalues ($<$300-400 MeV) are
excluded, so it forms an ideal complement to the Lanczos treatment
of the low eigenvalues. In the Monte Carlo simulations discussed in
this paper, we have used the pure Wilson plaquette gauge action
for the heat-bath sweeps and carried out the accept-reject test
using the truncated low-eigenvalue contribution to the determinant.
In the context of the more general Wilson-loop description of the
high-eigenvalue segment, our present procedure is equivalent to
approximating the high-eigenvalue contribution to  $\ln {\rm Det}\;H$
by just the sum of a constant and a plaquette term. (This is in the
same sense that the ordinary quenched approximation is equivalent
to approximating the {\it full} determinant by a shift of $\beta$.)

Perhaps the best news to emerge from these numerical simulations is that
the Metropolis test on the low-eigenvalue-truncated determinant
yields a reasonably large acceptance ratio after one or more complete
heat bath sweeps, even when the quarks are very light. This is in
marked contrast to what would happen if one tried to include the
full determinant via an accept-reject step between quenched Monte
Carlo sweeps. Even for a single heat bath sweep, the fluctuations
of the full determinant are much too large to yield a reasonable
acceptance rate. The Lanczos calculation of the lowest few hundred
eigenvalues requires an amount of computing time of the same order
as that of an ordinary conjugate-gradient inversion of the Dirac
operator. Thus, even if the accept-reject step is performed after
every sweep, the computing required is still comparable to that of other
full QCD algorithms such as hybrid Monte Carlo. Moreover, the 
performance of the algorithm does not seriously degrade in the
light quark limit, which may provide a significant advantage over
hybrid Monte Carlo for the study of chiral behavior in full QCD.
Finally, for issues associated with chiral symmetry, the
special handling of low eigenvalues is theoretically appropriate,
and the eigenmodes extracted by the Lanczos analysis provide
a detailed view of the connection between chiral symmetry 
breaking and the low-lying Dirac spectrum. 

\newpage
{\noindent \Large \bf Acknowledgements}

A. Duncan is grateful for the
hospitality of the Fermilab Theory Group, where this work was performed.
The work of A. Duncan was supported in part by NSF grant PHY97-22097.
The work of  E. Eichten was performed at the Fermi National
Accelerator Laboratory, which is operated by University Research Association,
Inc., under contract DE-AC02-76CHO3000. 
The work of H. Thacker was supported in part by the Department of Energy
under grant DE-AS05-89ER40518. Much of the numerical work was performed on the 
Fermilab ACPMAPS system.




\newpage

\begin{thebibliography}{99}
\bibitem{mqa}  W. Bardeen, A. Duncan, E. Eichten, G. Hockney and H. Thacker
Phys. Rev. {\bf } (1998).
\bibitem{qed2}  W. Bardeen, A. Duncan, E. Eichten, and H. Thacker,
Phys. Rev. {\bf } (1998).
\bibitem{qnchart} W. Bardeen, A. Duncan, E. Eichten, and H. Thacker, hep-lat 9806002.
\bibitem{coleman} ``The Uses of Instantons", in {\em The Aspects of Symmetry,
Selected Erice Lectures of S. Coleman}, (Cambridge 1985).
\bibitem{hermitian} R. ~Setoodeh, C.T.H. ~Davies, and I.M. ~Barbour, Phys. Lett {\bf B213} 195 (1988); K.M. ~Bitar, A.D. ~Kennedy, and P. ~Rossi, Phys. Lett. {\bf B234} 333 (1990).
\bibitem{LeutSmil} H. Leutwyler and A. Smilga, Phys. Rev. {\bf D46}, 5607 (1992).
\bibitem{overlap} R. Narayanan and H. Neuberger, Nucl. Phys. {\bf B443} 305 (1995).
\bibitem{chirlog} A. Morel, J. Phys (paris) 48,(1987)1111; S. Sharpe, Phys. Rev. D41,
(1990) 3233; C. Bernard and M. Golterman, Phys. Rev. D46 (1992)853.
\bibitem{multiboson} M. L\"{u}scher, Nucl. Phys. {\bf B418}, 637 (1994).
\bibitem{SmitVink} J. Smit and J. Vink, Nucl. Phys. B286(1987)485; J. Vink, Nucl. Phys. B307(1988)549.
\bibitem{Lat97} A. Duncan (speaker), W. Bardeen, E. Eichten, and H. Thacker, talk given at Lattice 97 (Edinburgh), in Nucl. Phys. Proc. Suppl. 63, 811 (1998).
\bibitem{conjgrad} T. Kalkreuter and H. Simma, Comput.Phys.Commun. {\bf 93} 33 (1996).
\bibitem{golubbook} G.H. Golub and C.F. Loan, {\em Matrix Computations}, 2nd edition
(Johns Hopkins, 1990).
\bibitem{cullwill} J. Cullum and R.A. Willoughby, J. Comp. Phys. {\bf 44} 329 (1981).
\bibitem{Kalk} T. Kalkreuter, Comput.Phys.Commun. {\bf 95} 1 (1996).
\bibitem{fbpaper} A. Duncan, E. Eichten, J. Flynn, B. Hill, G. Hockney, and H. Thacker,
Phys. Rev. {\bf D51} 5101 (1995).
\bibitem{deGrand} T. deGrand and A. Hasenfratz, Phys. Rev. {\bf D49} 466 (1994).
\bibitem{numrec} W.H. Press, S.A. Teukolsky, W.T. Vetterling and
 B.P. Flannery, {\em Numerical Recipes in C}, 2nd ed. (Cambridge University Press 1992).
\bibitem{Wein97} J.~C.~Sexton and D.~H.~Weingarten, Phys. Rev. {\bf D55}  4025 (1997).
\bibitem{Irving97} A.~C.~Irving and J.~C.~Sexton, Phys. Rev. {\bf D55}  5456 (1997);
A.C. Irving, J.C. Sexton and E. Cahill,hep-lat/9708004.
\bibitem{improvact} M. L\"{u}scher and P. Weisz, Phys. Lett. {\bf 158B} 250 (1985), and
 references therein.
\end{thebibliography}
\end{document}